\documentstyle[12pt,aasms4]{article}

\lefthead{Antiochos, DeVore, \& Klimchuk}
\righthead{Model of Mass Ejections}

\begin{document}

\title{The Magnetic Topology of Solar Eruptions}

\author{S. K. Antiochos}
\affil{E.O. Hulburt Center for Space Research, Naval Research Laboratory,
Washington, DC, 20375}

\begin{abstract}
We present an explanation for the well-known observation that
complexity of the solar magnetic field is a necessary ingredient for
strong activity such as large eruptive flares. Our model starts with
the standard picture for the energy build up -- highly-sheared,
newly-emerged magnetic field near the photospheric neutral line
held down by overlying unsheared field. Previously, we proposed
the key new idea that magnetic reconnection between the unsheared
field and neighboring flux systems decreases the amount of overlying
field and, thereby, allows the low-lying sheared flux to
``break out'' (\cite{Antiochos98}).  In this paper we show that a
bipolar active region does not have the necessary complexity for this
process to occur, but a delta sunspot has the right
topology for magnetic breakout. We discuss the implications of these
results for observations from SOHO and TRACE.
\end{abstract}


\section{Introduction}
One of the outstanding questions in solar physics concerns the
magnetic topology responsible for major eruptive activity such as
coronal mass ejections (CME) and large two-ribbon flares accompanied
by a filament/prominence eruption. This question is critical both for
understanding the physics of solar eruptions and for predicting
them. It has long been known empirically that two features are
required for strong solar activity: magnetic shear and magnetic
complexity (e.g., \cite{Patty86}). By shear we mean that the field
near the photospheric neutral line is observed to be almost parallel
to the neutral line, rather than perpendicular to it, as would be the
case for a potential field. By complexity we mean that that the active
region is not simply bipolar, consisting of a leading sunspot and a
well-separated trailing polarity region, but contains instead an
intermingling of polarities as in a delta sunspot in which two
opposite polarity umbrae share a common penumbra (\cite{Bray64}).
Delta sunspots, in particular, are well known to be prolific flare
producers. (e.g., \cite{Patty86,Tanaka91,Zhang95}).

The physical reason for the observed requirement of shear is
straightforward to understand.  The field must contain free energy in
order to power the eruptive event; somewhere, there must be a large
deviation of the field from its potential state.  Furthermore, since
the free energy is most easily generated in the deep photosphere, it
should appear in the field that has most recently emerged and,
therefore, is nearest the neutral line.

The second feature, magnetic complexity, is more difficult to
understand. CMEs, filament eruptions, and two-ribbon flares usually
appear to occur along only a single neutral line, i.e. in a bipolar
region. The post-eruption X-ray emission is observed to straddle only
one neutral line. In fact, the standard picture for flare/filament
eruptions, the so-called tether cutting model
(e.g. \cite{Sturrock89,Moore92}) involves only a single sheared
magnetic arcade.  Although magnetic complexity plays no role in this
model, the observations show that it is essential for the event to
occur in the first place.  Also, recent observations indicate that
multiple neutral lines may be a common feature of CMEs
(\cite{Webb97}).  In this paper, we clarify why the magnetic
complexity of a delta sunspot, or worse, is necessary for violent
eruptions.

\section{Magnetic Topologies}
Recently, we developed a new model for solar mass ejections
(\cite{Antiochos98}) which shares with the standard picture, the
features that the eruption and the bulk of the energy release occur in
a sheared arcade. As usual, the shear is concentrated near the neutral
line, so that the stressed core field near the neutral line provides
the free energy and the upward push, whereas the unsheared overlying
field provides the downward pull that restrains expansion of the core
field. The key new feature proposed in our model is that additional
flux systems are present, which make it possible for reconnection to
take place between the unsheared overlying flux and the flux in those
neighboring systems.  This reconnection transfers unsheared flux to
the neighboring flux systems, thereby removing the overlying field and
the restraining pull. Hence, reconnection allows the innermost core
field to ``break out'' to infinity without opening the overlying field,
and violating the Aly-Sturrock open-field energy limit
(\cite{Aly91,Sturrock91}).

A 2D model for this ``breakout'' process has been presented
(\cite{Antiochos98}). The model postulated a global magnetic field
topology for the Sun consisting of a four-flux system in which
reconnection between the upper and lower systems transfers flux to the
two side systems. The question we address in this paper is the
following: {\it What is the minimum complexity needed in the magnetic
field of an active region so that a similar process can occur in a
fully 3D geometry?}

In order to answer this question, let us first determine the topology
of a bipolar active region. For this case, the relevant magnetic field
corresponds to a large-scale background solar bipole in which an
active region bipole has emerged.  Since we are interested only in the
basic topology, it is sufficient to consider a potential, unsheared
field; a typical example is shown in Figure 1.  The yellow surface in
the Figure corresponds to the photosphere chosen to be at $z=0$, and
the view is oriented so that the $+z$ direction is vertical, $+y$ is
to the right, and $+x$ is out of the page.  The background field is
that due to a magnetic dipole pointing in the $-y$ direction, with
relative magnitude $10^6$, and located at position $(0,50,-100)$. The
active region is due to another dipole, in the $+y$
direction, with magnitude $2\times 10^4$, and located at
$(0,12,-10)$. The active region ``sunspots'' are indicated in the
Figure by the black and white spots, which contain all the contours of
$B_z$ on the photosphere that are greater than half-maximum. Note that
$B_z$ is smooth everywhere; the contours simply indicate where there
is concentration of field, and can be thought of as indicating the
umbra of the spots. A true sunspot would have much higher field
concentration than our example here, but the topology would be the
same.

There are three polarity regions on the photosphere: a semi-infinite
negative region due to the background dipole, a semi-infinite positive
region consisting of the positive background plus the positive,
leading-polarity active region sunspot, and a finite negative region
due to the trailing polarity sunspot. The two neutral lines defining
the three polarity regions are indicated by the black lines on the
yellow photospheric plane. The closed neutral line surrounding the
finite negative polarity region would correspond to the edge of the
trailing spot's penumbra.

It is evident from Figure 1 that the topology of a bipolar active
region is simply that of an embedded dipole (\cite{Antiochos96}). It
consists of only two flux systems, the flux connecting the positive
polarity region to the background field (red field lines) and the flux
connecting the positive polarity to the negative spot (green
lines). The separatrix surface can be visualized by examining those
field line sections where the four pairs of red and green lines are
indistinguishably close together in the Figure. These red-green
sections outline a hemispherical separatrix surface that encloses the
green flux, and across which the magnetic connectivity is
discontinuous.  The connectivity is also discontinuous at a singular
line originating at a point inside the negative sunspot region (where
the four green field lines appear to come together in the Figure), and
continuing into the background region (the four red field lines that
are close together).  A magnetic null occurs in the corona where this
singular line passes through the separatrix surface
(e.g. \cite{Lau90}).  The intersection of the separatrix surface with
the plane of the photosphere forms a closed curve that encircles the
negative sunspot and defines the photospheric boundary between the
green and red field lines. The photospheric connectivity is
discontinuous at this curve and at the two points where the singular
line intersects the photosphere.

We now imagine that the field near the photospheric neutral line of
the negative spot is strongly sheared, and ask what happens to the
overlying green flux. The key point is that, in this case, the amount
of overlying unsheared flux cannot decrease.  The green flux inside
the photospheric separatrix curve must always be exactly equal to the
amount of flux in the trailing negative-polarity sunspot.  Note,
however, that reconnection and energy release can still occur.  For
example, if the ``sunspots'' in Figure 1 move, then red and green flux
can reconnect at the null (\cite{Antiochos96}), exchanging positions
and causing the separatrix curve to move to its most energetically
favorable position.  But the amount of green flux inside the
separatrix never changes. Therefore, in order for the sheared core
field to open, a fixed amount of unsheared overlying flux would have
to open as well, which is energetically unfavorable
(\cite{Aly91,Sturrock91}). This is the fundamental reason why bipolar
active regions do not exhibit violent eruptive activity even if
strongly sheared -- their topology is too restrictive to allow a
decrease in the overlying unsheared flux.

Now, let us consider a delta sunspot. To the system of Figure 1, we
add a dipole that is pointing in the $+z$ direction, with magnitude $7
\times 10^4$, and located inside the negative spot region at
$(0,0,-6.5)$.  The resulting topology is shown in Figure 2, which
focuses on the region around the negative spot for clarity. Due to the
presence of the additional dipole, a positive polarity umbra and a
neutral line have appeared inside the the negative spot, giving it a
delta appearance. In order to make the topology easier to visualize,
we chose the dipole parameters so that the parasitic polarity is
located centrally within the parent spot, but the results described
below remain valid for the usual case where the parasitic spot is off
to one side.

Now four distinct flux systems are present. In addition to the
original red and green systems, there is a blue flux system comprised
of field lines that connect the negative spot to the parasitic
positive polarity. We expect this system to appear with the addition
of the parasitic polarity, but surprisingly, another flux system (gray
field lines) appears inside the parasitic spot and connects some of
the innermost positive flux to the distant, negative-polarity
background region.  The separatrix surfaces now consist of a
horizontal and a vertical torus, both cut by the photosphere plane.
The horizontal torus can be visualized by joining the fieldline
sections where the red and green lines are indistinguishably close to
the sections where the blue and gray lines are close. Similarily the
vertical torus is formed by joining the blue-green sections to the
red-gray sections, (only a small base section of this torus can be
seen in the Figure).  In the corona, the curve of intersection of the
two torii defines a separator line along which all four flux systems
come in contact. For the particular delta spot of Figure 2, which has
a high degree of symmetry, there are four magnetic nulls located on
this separator line, but there can be as few as two.  The intersection
of the two separatrix surfaces with the plane of the photosphere forms
three concentric closed curves across which the photospheric
connectivity is discontinuous. As before, an approximately circular
curve outside the negative spot marks the photospheric boundary
between red and green flux, a closed curve lying inside the negative
spot but outside the positive spot denotes the boundary between green
and blue, and a closed curve lying inside the positive parasitic spot
bounds the blue from the gray.

This four-flux topology is precisely what is needed for eruption to
occur, and is the direct 3D analog of our 2D model
(\cite{Antiochos98}).  Assume, as before, that the neutral line around
the negative spot is strongly sheared. The green flux overlying the
core neutral line field can now be decreased by reconnection between
the green and gray systems at the coronal separator line. This will
convert green and gray flux into red and blue. Note that, unlike the
dipolar active-region case above, the actual amount of green flux
decreases, which allows the sheared core field to erupt outward.  On
the photosphere, the outer separatrix curve shrinks inward while the
middle separatrix curve expands outward. An identical behavior can be
obtained by shearing the inner neutral line. Reconnection between blue
and red field lines at the separator line converts them into green and
gray. Of course, a real active region on the Sun can have much more
complexity than this very simplest of delta-configurations; even so,
we expect that the topology of four flux systems meeting along a
coronal separator line is the basic topology underlying eruptive
activity. It is unlikely that more than four systems would share a
common boundary.

We conclude, therefore, that a delta sunspot has sufficient complexity
for the breakout model to operate, and propose that this is why delta
spots are so active. Note that, if eruption does occur, it will take
place on only one neutral line at a time, because a decrease in the
field overlying one neutral line requires an increase in the field
overlying the adjacent neutral line.  This explains why solar
eruptions typically appear to involve only one sheared arcade.

The eruption, itself, is exactly the same in our model as in the
standard picture, except that now even the lowest-lying sheared field
can open (\cite{Antiochos98}). The sheared field at one neutral line
blows open, ejecting a filament if present, and then reconnects with
itself to close back down and form the observed hot x-ray loops.  It
should be emphasized that this post-eruption reconnection, which can
release a great deal of energy, is completely different than the
reconnection proposed in our model, which acts only as a trigger for
the eruption.

Our model has interesting implications for observations.  A crucial
feature of the topology in Figure 2 is the presence of the gray flux
connecting the parasitic spot to the distant background region.  This
feature is the main difference between the delta and the bipolar field
of Figure 1, in which all the flux of the negative sunspot connects
locally.  Even though the parasitic-polarity flux in our example is
significantly smaller than the flux of the parent spot, this nonlocal
connection is the minimum-energy, current-free state of the magnetic
field. Of course, if the parasitic spot is sufficiently small, then
all its flux will close locally into the parent spot, so that the
topology becomes that of an embedded dipole within an embedded
dipole. These considerations imply that, as flux emerges through the
photosphere, one should see an abrupt jump in the structure of coronal
loops after it becomes energetically favorable for the parasitic spot
to form distant connections. Such dramatic jumps in coronal connections,
correlated with magnetic emergence, should be directly observable with
the magnetogaph on SOHO and the high-cadence XUV telescopes on TRACE.

\acknowledgments

This work has been supported in part by NASA and ONR.

\clearpage

%
%

\vfill\eject

\begin{figure}
\caption{ The magnetic topology of a dipolar active region. \label{fig1}}
\end{figure}

\begin{figure}
\caption{ The magnetic topology of a delta sunspot region. \label{fig2}}
\end{figure}

\end{document}